\documentstyle[11pt,epsfig,emulateapj]{article}
\input epsf
\newcommand\kms{$\rm km\;s^{-1}$}
\newcommand\percc{$\rm cm^{-3}$}
\newcommand\percmsq{$\rm cm^{-2}$}
\newcommand\HII{\ion{H}{2}}

\newcommand\CI{\ion{C}{1}}
\newcommand\thCI{$^{13}$\ion{C}{1}}
\newcommand\thC{$^{13}$C}
\newcommand\mthC{{\rm ^{13}C}}
\newcommand\twCI{$^{12}$\ion{C}{1}}
\newcommand\twC{$^{12}$C}
\newcommand\mtwC{{\rm ^{12}C}}

\newcommand\Cplus{{$\rm C^+$}}
\newcommand\twCplus{{$\rm ^{12}C^+$}}

\newcommand\thCplus{{$\rm ^{13}C^+$}}

\newcommand\thPto{{${^3P_2} - {^3P_1}$}}
\newcommand\thPoz{{${^3P_1} - {^3P_0}$}}
\newcommand\Ffivehalves{{$F=\slantfrac{5}{2}- \slantfrac{3}{2}$}}
\newcommand\twCO{{$\rm ^{12}CO$}}
\newcommand\thCO{{$\rm ^{13}CO$}}
\newcommand\CeiO{{$\rm C^{18}O$}}
\newcommand\twCeiO{{$\rm ^{12}C^{18}O$}}
\newcommand\thCeiO{{$\rm ^{13}C^{18}O$}}
\newcommand\mCeiO{{\rm C^{18}O}}

\newcommand\mthCeiO{{\rm ^{13}C^{18}O}}

\lefthead{Keene et al.}
\righthead{Submm Detection of \thCI\ in the ISM}

\submitted{Astrophysical Journal Letters, in press}

\begin{document}

\title{Detection of the ${^3P_2} \to {^3P_1}$ Submillimeter
  Transition of $^{13}$CI in the\\
  Interstellar Medium: Implication for Chemical Fractionation}

\author{Jocelyn Keene,\altaffilmark{1} Peter Schilke,\altaffilmark{2}
  J. Kooi,\altaffilmark{1} D. C. Lis,\altaffilmark{1}\\ David M.
  Mehringer,\altaffilmark{1} and T. G. Phillips\altaffilmark{1} }
\altaffiltext{1}{Caltech Submillimeter Observatory, Caltech 320-47,
  Pasadena, CA 91125; jbk,kooi,dcl,mehring,phillips@tacos.caltech.edu}
\altaffiltext{2}{Max-Planck-Institut f\"ur Radioastronomie, Auf dem
  H\"ugel 69, 53121 Bonn, Germany; schilke@mpifr-bonn.mpg.de}
\authoremail{jbk@tacos.caltech.edu}

\begin{abstract}
  We report the first detection of the submm emission from the \thC\ 
  isotope of atomic carbon in the ISM.  The \Ffivehalves\ component of
  the \thPto\ transition was observed with the CSO in a region $\sim
  4$\arcmin\ S of Orion IRc2, near the western end of the Orion Bar.
  The \twC\ to \thC\ isotopic abundance ratio is $58\pm 12$ corrected
  for opacity of the \twCI\ line and the fractional intensity of the
  \thCI\ hyperfine component (60\%).  This is in agreement with the
  value for the equivalent ratio in \Cplus. In comparison, our
  measurement of the \CeiO\ to $^{13}$\CeiO\ ratio from observations
  of 2--1 and 3--2 lines toward the same position gives a value of
  $75\pm 9$.  PDR models predict that the \twC\ to \thC\ abundance
  ratio is particularly sensitive to chemical fractionation effects.
  If $\rm ^{13}C^+$ is preferentially incorporated into \thCO\ at
  cloud edges there will be a dramatic reduction in the abundance of
  \thC. This is contrary to our observations, implying that the
  importance of chemical fractionation is small or is compensated for
  by isotopic-selective photo-dissociation of \thCO\ in this region
  with a large UV illumination.
  
\end{abstract}


\section{Introduction}

The abundances and ratios of the CNO isotopes in the interstellar
medium (ISM) are sensitive to details of stellar and Galactic chemical
evolution.  
Many attempts have been made to measure the ratio of \twC\ to \thC\ 
throughout the Galaxy and in the local ISM, as well as in stars, to
help guide the Galactic evolutionary models.  Optical, UV, and IR
absorption line techniques have been used for these measurements as
well as radio emission line measurements (see the review by Wilson \&
Rood 1994).

One of the most successful attempts for ISM measurements has been made
by Langer \& Penzias (1990 \& 1993; hereafter LP90 \& LP93) who
compared the weak emission lines of \CeiO\ and \thCeiO.  They found a
Galactic gradient in the \twC\ to \thC\ ratio ranging from $\sim 25$
near the Galactic center to about 60 to 70 at a Galactic radius
($R_G$) of about 10 kpc.  The value in the local ISM is $62\pm 4$.
For $R_G > 16$ kpc, Wouterloot \& Brand (1996), using the same
technique, found a value $>100$.  Ratios derived from observations of
$\rm H_2CO$ show a similar Galactic gradient but are $\sim 50$\%
higher (Wilson \& Rood 1994).  Unfortunately, ratios derived from both
CO and $\rm H_2CO$ can be distorted by chemical fractionation and
isotopic-selective photo-dissociation, two processes discussed further
in \S 4.

Centuri\'on et al.\ (1995) have reviewed the various determinations
using optical absorption lines of $\rm CH^+$ and concluded that the
abundance ratio in the local ISM is $67\pm 3$.  Boreiko and Betz
(1996) have measured the \twCplus\ to \thCplus\ ratio for the average
of several positions toward the Orion Nebula region.  The value they
found is $58\pm 6$. It is thought that neither of these probes should
suffer from chemical fractionation effects.  

In this letter we describe measurements of the ratio with another
fundamental probe, neutral atomic carbon. According to PDR models, the
ratio of atomic \twC\ to \thC\ in the dense interstellar medium is
particularly sensitive to the opposing effects of chemical
fractionation and isotopic-selective photo-dissociation which can also
affect the isotopomeric ratios of many molecules, e.g., CO and ${\rm
  H_2CO}$.  In this paper we use the atomic carbon isotopic ratio to
assess the relative importance of these effects in one astronomical
source. We compare the isotopic ratio in atomic carbon to that in CO,
through measurements of the weak lines of \thCeiO\ and \CeiO\ which
are similarly affected by chemical fractionation and
isotopic-selective photo-dissociation as the more abundant species
(Langer et al.\ 1984; van Dishoeck \& Black 1988).

The \thPoz\ transition of \twCI\ at 492 GHz has been available for
high spectral resolution measurements in molecular clouds for many
years.  Unfortunately, the equivalent \thCI\ transition is
insufficiently split from the \twCI\ line to allow its discrimination;
the largest split is only 3.6 MHz (2.2 \kms; Yamamoto \& Saito 1991).
However, the splitting between the \twCI\ (\thPto) and the strongest
hyperfine component of the \thCI\ equivalent transition, both lying
near 809 GHz, is 152 MHz (56 \kms; Klein et al. 1997), providing a
better opportunity for detection of \thCI.

\section{Observations}

The observations were made during two periods of good weather in 1996
December and 1997 February with the recently available 850 GHz receiver
(Kooi et al.\ 1997) at the Caltech Submillimeter Observatory (CSO).
The 225 GHz zenith opacity at the time was $\sim 0.03$, corresponding
to an 809 GHz opacity of $\sim 0.7$.  We observed \twCI\ and \thCI\ in
the upper sideband in 1996 December and in the lower sideband in 1997
February. There was excellent agreement between the two observing
runs. Using Mars (diameter 10\farcs 6), we measured the main beam
efficiency at 809 GHz to be 33\% in 1997 February assuming the
theoretical beamsize of 9\arcsec.  Our source is very extended, so the
Moon efficiency would be more appropriate to use in correcting the
measured intensities. We estimate it to be $60\% \pm 20\%$.  In
addition we measured some of the lines of CO and its isotopomers in
the 2--1 and 3--2 transitions.

We decided to observe the \Ffivehalves\ component of the \thPto\ 
transition of \thCI\ because it is the strongest of the three
components, with 60\% of the total line-strength.  We centered this
component within the quietest part of our bandpass which meant that
the two weaker components were not observed.

We chose to observe a position in the clump at the base of the Orion
Bar ionization front, $\alpha(1950) = \rm 5^h 32^m 47\fs 7$,
$\delta(1950) = -5\arcdeg 28\arcmin 30\arcsec$, ($\sim 4\arcmin$ S of
IRc2) because it is known to have bright 492 GHz \CI\ lines (Tauber et
al.\ 1995).  Unlike the Bar itself, the clump does not appear to have
a strong edge-on geometry. It also has a simple molecular spectrum,
thus avoiding line confusion such as occurs near Orion IRc2.

The fine-structure energy level intervals of the \twC\ and \thC\
atoms have been measured with high precision in the laboratory by
Saykally \& Evenson (1980), Cooksy et al.\ (1986), Yamamoto \& Saito
(1991), and most recently by Klein et al.\ (1997, Letter this issue).

\section{Results}

The observed spectra of \thCI\ and \twCI\ and the comparison
isotopomeric lines of CO are shown in Figure 1.  This is the first
detection of submillimeter emission from \thC\ in the ISM. We measured
the frequency of the \twCI\ and \thCI\ lines by comparison with the
\thCO\ (2--1) line (not shown) which has a shape nearly identical to
that of the \twCI\ line. The frequency we measure for the \thPto\ 
transition of \twCI\ is $809342.3 \pm 0.4$ MHz and the frequency of
the \Ffivehalves\ component of the \thPto\ transition of \thCI\ is
$809492.8 \pm 1.1$ MHz.  These agree well with the new laboratory
measurements by Klein et al.\ (1997).

To compare the atomic isotopic ratio to the molecular isotopomeric
ratio we need to derive the ratio of column densities from the ratio
of line intensities for both sets of data.

\subsection{Atomic Carbon Isotopic Ratio}

The column density ratio is derived from the line intensity ratio as
follows:

\begin{displaymath}
  \frac{N({\rm^{12}C})}{N({\rm ^{13}C})} = \frac{S({\rm ^{13}C})}
  {\beta({\rm ^{12}C})} \; \frac{I({\rm ^{12}C})}{I({\rm ^{13}C})},
\end{displaymath}

\noindent
where $N$ indicates column density, $I$ is the measured integrated
line intensity (in K \kms), $S$ is the fractional strength of the
\Ffivehalves\ component of the \thPto\ transition of \thCI, and $\beta
= (1-e^{-\tau})/\tau$ is the escape probability of the \twCI\ line
photons, which compensates for the finite optical depth, $\tau$, in
the \twCI\ line.

Because of the low signal-to-noise ratio of the weak \thCI\ line, we
use three methods of measuring line intensity ratios. First, we
integrate the line intensities over the interval 5 to 14 \kms.  This
method has the disadvantage of including excess noise outside the line
core. Next, we integrate over the \thCI\ and \twCI\ line profiles
weighted by the \twCI\ line profile, a method used by LP90 for CO
isotopomers to lessen the contribution of noise outside the line core.
Finally, we fit Gaussians to the \twCI\ line and use the same width
and central velocities for the \thCI\ line.
The dispersion in these determinations is low (4\%) and we adopt the
mean value, 80.  The S/N in the line integral measurement for the
\thCI\ line estimated from the noise in the baseline is $\sim 6.5$ so
we assign a 15\% uncertainty to the ratio. Thus the measured line
intensity ratio is $80\pm 12$.  Because the \thCI\ and \twCI\ lines
were measured simultaneously in the same receiver sideband, there is
no significant calibration uncertainty in the line intensity ratio.

The \thCI\ transition has 3 hyperfine components.  We assume that the
excitation is thermal so that the hyperfine line intensities are
proportional to the quantum mechanical line strengths and $S(\mthC) =
0.6$.

\epsfysize=6.8in\epsfbox[115 70 500 640]{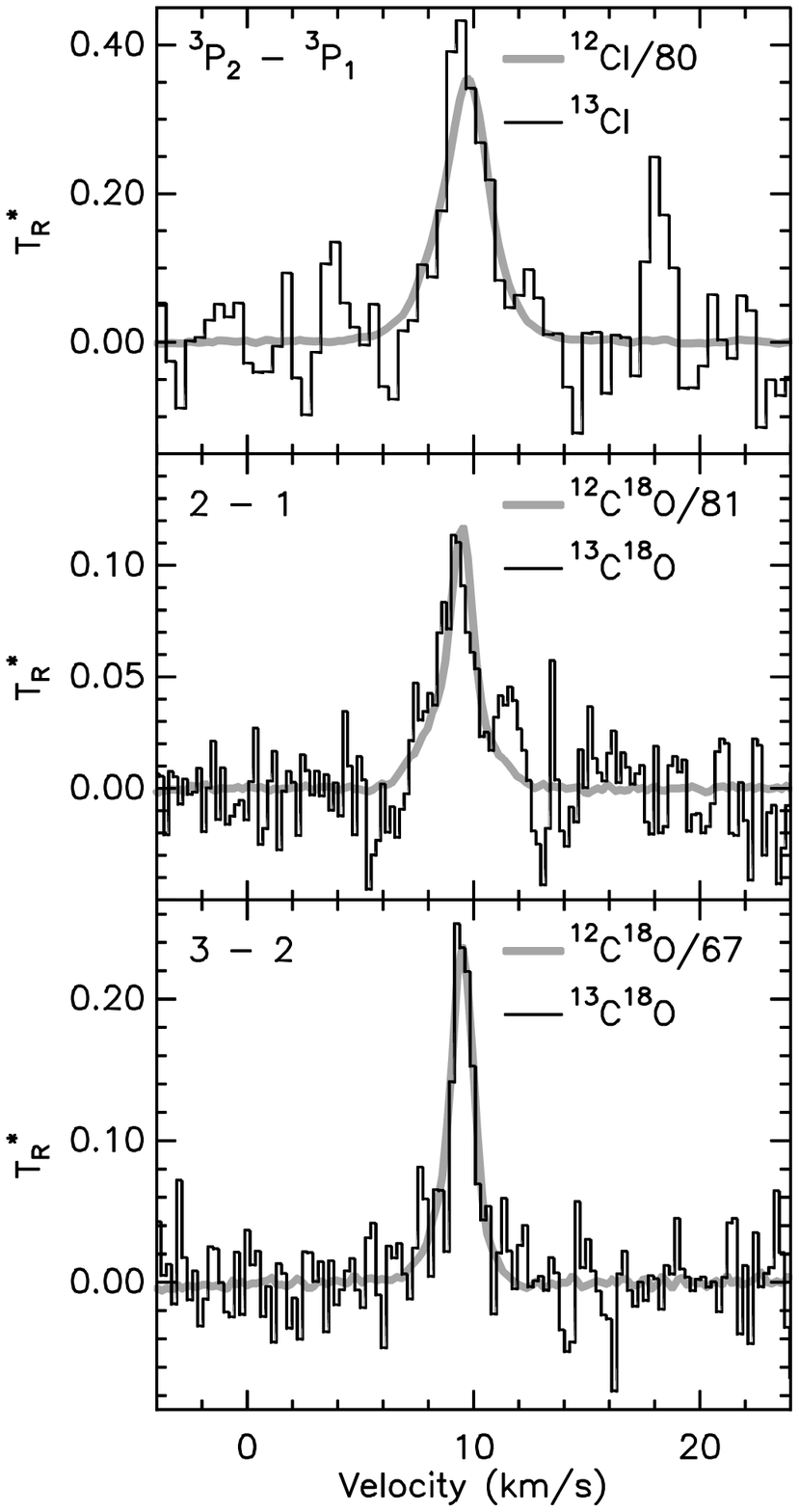}
\figcaption[keene_figure1.eps]{\small Observed spectra of ({\it top}) the
  \protect\thPto\ transitions of \protect\twCI\ and \protect\thCI\ 
	(\protect\Ffivehalves); ({\it
    middle}) the 2--1 transitions of \protect\twCeiO\ and \protect\thCeiO; 
    ({\it
    bottom}) the 3--2 transitions of \protect\twCeiO\ and \protect\thCeiO.
  All spectra
  have been corrected for the estimated Moon efficiencies, 60\% for
  \protect\CI\ and 70\% for CO.  The spectra of the species containing
 \protect\twC\ 
  have been additionally scaled as indicated. In creating the velocity
  scales for these plots the frequencies for the CO isotopomers have
  been taken from Winnewisser, Winnewisser, \& Winnewisser
  (1985).\label{spectra}}
\vspace{0.3in}

The most difficult factor to estimate is the correction for the finite
optical depth of the \twCI\ line. We assume that the \CI\ excitation
temperature is 104 K, as we measured for the \twCO\ (2--1) line, and
that the telescope beam efficiency for large sources is $60\% \pm
20\%$. The uncorrected line intensity of 17 K then corresponds to a
true line intensity of $28.3^{+14}_{-7}$ K and an opacity $\tau =
0.40^{+0.28}_{-0.12}$ with line escape probability $\beta =
0.83^{-0.11}_{+0.04}$.

The resulting ratio of the \twC\ to \thC\ column densities is $58\pm
12$, where we have added the uncertainty in the escape probability in
quadrature with that of the line intensity ratio.

\subsection{CO Isotopomeric Ratios}

The CO isotopomeric column density ratio is derived from the line
intensity ratio as follows:

\begin{displaymath}
\frac{N({\mCeiO})}{N({\mthCeiO})} = \frac{F}{\beta({\mCeiO})}\;
\frac{I({\mCeiO})}{I({\mthCeiO})},
\end{displaymath}

\noindent
where $F$ is a small correction due to the difference in the molecular
excitation and the transition probabilities caused by the difference in
frequency between the two lines.

We use the same three methods described in \S 3.1 for determining
line intensity ratios and adopt
the mean values of these determinations: 81 for the 2--1 lines and 67
for 3--2.  Because the \thCeiO\ lines are very weak, the uncertainties
in the line ratios are substantial. One measure of the uncertainties
is the ranges in the ratios determined by the three different methods.
These are $\sim 7$\% for the 2--1 ratio and $\sim 12$\% for 3--2.  The
uncertainties in the integrated line intensities for the \thCeiO\ 2--1
and 3--2 lines, as estimated from the noise in the baselines, are 10\%
and 9\% respectively. To be conservative we adopt the larger of the
two uncertainty estimates. We have estimated the telescope efficiency
to be 70\% for both 2--1 and 3--2 transitions based on past
determinations.  Because the frequencies of the two lines in each
frequency range are so close, this should add little uncertainty to
the ratios.  A larger factor is the uncertainty in the sideband ratio,
since we have double sideband receivers. We estimate this uncertainty
to be 10\% for each measurement or 14\% for the ratio.  Added in
quadrature to the previous uncertainties, we have 17\% uncertainty for
the 2--1 line intensity ratio and 19\% for 3--2.

Using standard LTE column density formulas we find for both the 2--1
and 3--2 lines that $F=0.92$. We can also estimate this correction
factor using an LVG calculation. Then we find $F=0.91$ for 2--1 and
$F=0.94$ for 3--2, so uncertainty in determination of $F$ is small
and can be neglected.

We have estimated the optical depth of the \CeiO\ lines using an LVG
model for the \CeiO\ (2--1) and (3--2) line intensities and ratio. The
temperature and density are both constrained on the low ends by the
line ratio which gives $T > 70$ K and $n({\rm H_2}) > 1.6\times
10^4$ \percc.  We adopt the best fit for the temperature given by the
brightness temperature of the CO (2--1) line ($T = 104~{\rm K},~
n({\rm H_2}) = 3\times 10^4$ \percc) which has opacities of 0.09 and
0.27 and escape probabilities of 0.96 and 0.88 for the 2--1 and 3--2
lines respectively.  These lie near the center of the possible range
as derived from the LVG model. As a measure of the uncertainties we
use the ranges of the escape probabilities, determined by the LVG
calculation with different densities and temperatures.  We find an
uncertainty of $\sim 3$\% for the 2--1 lines and $\sim 9$\% for the
3--2 lines.  The total uncertainty for $N({\mCeiO})/N({\mthCeiO})$ is
$\sim 17$\% for the 2--1 line and $\sim 21$\% for the 3--2 line.

Finally, combining all the factors, we find that
${N({\mCeiO})}/{N({\mthCeiO})} = 77\pm 12$ from the 2--1 lines and
$72\pm 15$ from 3--2, with a weighted average of $75 \pm 9$.
In comparison, values of the $\rm ^{12}C$ to $\rm ^{13}C$ ratio toward
the Orion Molecular Cloud previously derived from observations of
\CeiO\ and $^{13}$\CeiO\ (2--1) and (1--0) range from $79\pm 7$ near
IRc2 (LP90) to $63\pm 6$ at a position $\sim 14$\arcmin\ S of IRc2
(LP93). Our value lies within this range and closer to the IRc2 value
as befits its position close to the \HII\ region. The average for the
solar neighborhood is $62\pm 2$ (LP93), so the positions near the M42
\HII\ region in Orion appear to be truly different from the average
neighborhood value, possibly due to the strong UV field of the
Trapezium.

\section{Discussion}

As mentioned in \S 1 there are two major physical phenomena which can
modify the abundance ratios of the various atomic and molecular forms
of carbon and its isotopes. The first is chemical fractionation of
\thCO\ and \twCO, due to the rapid exothermic exchange of \thCplus\ 
with \twC\ in CO preferentially forming \thCO\ (Watson, Anicich \&
Huntress 1976).  Since \thC\ is formed by the recombination of
\thCplus, \thC\ will be depleted in favor of \thCO.  This process is
most effective where there is a large abundance of $\rm C^+$ ions and
where the temperature is low, i.e., in the transition zone between
ionized and molecular forms of carbon near the edges of molecular
clouds. Here ionized, atomic and molecular forms of carbon co-exist
and the temperature is reduced due to the enhanced cooling by atomic
and molecular species.  However, if the temperature were high the
rapid ion exchange would push the \twCO\ to \thCO\ ratio toward the
elemental ratio.

The second process is the isotopic-selective photo-dissociation of CO.
Again, this is generally most effective near the edges of molecular
clouds.  It occurs because CO and its isotopes are partially
self-shielding, in that the photo-dissociation of CO occurs through
discrete lines in the ultraviolet.  $\rm H_2$ molecules and dust act
to partially shield CO --- $\rm H_2$ through its heavily
saturation--broadened UV absorption lines and dust through continuum
UV absorption . \twCO\ is also very effective at shielding itself,
because of its large abundance, but \thCO\ is much less so, resulting
in a greater photo-dissociation rate for \thCO\ and enhancement of
\thC\ and \thCplus\ abundances at the expense of \thCO.  The situation
for photo-dissociation of \CeiO\ and \thCeiO\ seems to be similar,
although the overall rates are higher for the less abundant species
(van Dishoeck \& Black 1988).

We have used the photo-dissociation region (PDR) model of Le Bourlot
et al.\ (1993) to interpret our observations. This model incorporates
both chemical fractionation and isotopic-selective photo-dissociation
and we have used full radiative transfer, explicitly computing the
mutual- and self-shielding of $\rm H_2$ and all the CO isotopomers. We
ran this model for a cloud of constant density $n({\rm H_2}) =
2.5\times 10^4$ \percc\ with impinging UV field of strength
$G_o=4\times 10^4$ (Tauber et al.\ 1994) and abundances from Flower et
al (1995).  The total gas phase isotopic abundance ratios were taken
to be $\rm ^{12}C/^{13}C = 60$ and $\rm ^{16}O/^{18}O = 500$. The
results are shown in Figure 2.

Figures 2{\it a} \& {\it b} show abundances and column densities of
the majority carbon species as a function of extinction in the model
cloud. Figures 2{\it c} \& {\it d} show the abundance and column
density ratios of the \twC\ bearing species to their respective \thC\ 
bearing species, normalized by their input ratios. Although most of
the variation in the isotope ratios in 2{\it c} takes place at visual
extinctions between about 2 and 7, we have observed a region of high
column density, $A_V > 50$ as estimated from the \CeiO\ column density
marked in 2{\it b} ($2.5\times 10^{16}$ \percmsq), so that these
surface effects are seen integrated with the cloud interior, as shown
in 2{\it d} and are thus, except for the atomic \twC\ to \thC\ ratio,
greatly diluted.

An important feature of the model is that at extinctions between about
2 and 7 both \thCO\ and \thCeiO\ are enhanced through
chemical fractionation relative to their \twC\ isotopomers by factors
up to 2.5 (Fig. 2{\it c}).  The double ratio $n({\rm
  ^{12}C^{18}O})/n({\rm ^{13}C^{16}O})$ displays much more extreme
behavior. This ratio is affected in the same sense by both chemical
fractionation and isotopic-selective photo-dissociation.  The first
increases the abundance of \thCO\ while the second decreases the
abundance of \CeiO.  The result is that \thCO\ is enhanced relative to
\CeiO\ by a factor up to 20 in the same range of extinction.

Another important feature is that the local abundance ratio of $n({\rm
  ^{12}C}) /n({\rm ^{13}C})$ closely follows that found in C$^+$ (Fig
2{\it c}).  These ratios vary in the opposite sense to the ratios found
in CO. In this model, {\it where the temperature is low in the PDR ($\sim
25$ {\rm K}), chemical fractionation dominates over isotopic-selective
  photo-dissociation} so that $^{13}$C is underabundant in both atomic
carbon and C$^+$ but is overabundant in CO.

Finally, although the {\it local abundance ratios} of C and C$^+$
track each other very well (Fig. 2{\it c}), there is no effect due to
the PDR visible in the {\it column density ratio in} C$^+$ (Fig. 2{\it
  d}), because C$^+$ is most abundant outside the molecular cloud
where it is a majority species and where this effect is negligible.
Therefore, according to this model, {\it the isotopic column density
  ratio measured in {\rm C$^+$} should give the true isotopic gas
  phase abundance ratio}. Boreiko and Betz (1996) have measured the
\twCplus\ to \thCplus\ ratio for the average of several positions in
the Orion Nebula region.  The value they find for the ratio is $58\pm
6$. In contrast, {\it atomic carbon is most abundant in the model
  where the isotopic fractionation effects are most important}.
Therefore there should be a large effect on the isotopic column
density ratio of atomic carbon, as seen in the model (Fig. 2{\it d}).
Yet the value we measure for the \twC\ to \thC\ ratio is $58 \pm 12$,
the same as the ratio in $\rm C^+$.

Our model in Figure 2{\it d} does not provide a good match to the
observations indicated. The observed ratio of \twC\ to \thC\ is the
same as the intrinsic ratio measured in $\rm C^+$ by Boreiko \& Betz
(1996) and is lower than that in CO as measured using the rare
isotopomeric species.  The deviation is in the sense that this model
evidently has a great deal too much chemical fractionation, which
pushes the model ratio of \twC\ to \thC\ to a larger value than is
consistent with our observations.

We were able to achieve better agreement between our data and the
model by artificially increasing the temperature within the model
cloud.  This higher temperature suppresses chemical fractionation
within the PDR. That such a high temperature ($\sim 120$ K) may be
present is indicated from observations of \thCO\ (6--5) toward the
Orion Bar (Lis, Schilke \& Keene 1997) but how it can be achieved
physically is beyond the scope of this Letter to discuss.

In summary, we have measured the \twC\ to \thC\ ratio in atomic carbon
($58\pm12$) and find it to be the same as that in ionized carbon
($58\pm 6$; Boreiko \& Betz 1996) and probably smaller than that found
in \CeiO\ ($75\pm 9$). The ratio of \twCplus\ to \thCplus\ should
 indicate the intrinsic isotopic abundance ratio since this is a
majority species outside molecular clouds and it should not be
significantly influenced by chemical fractionation or
isotopic-selective photo-dissociation at cloud edges.  From the PDR
model, we see that the agreement between the isotopic ratios in
ionized and atomic carbon implies that in this highly illuminated
region chemical fractionation must be unimportant or is compensated
for by isotopic-selective photo-dissociation.

\epsfysize=6.4in\epsfbox[120 40 500 700]{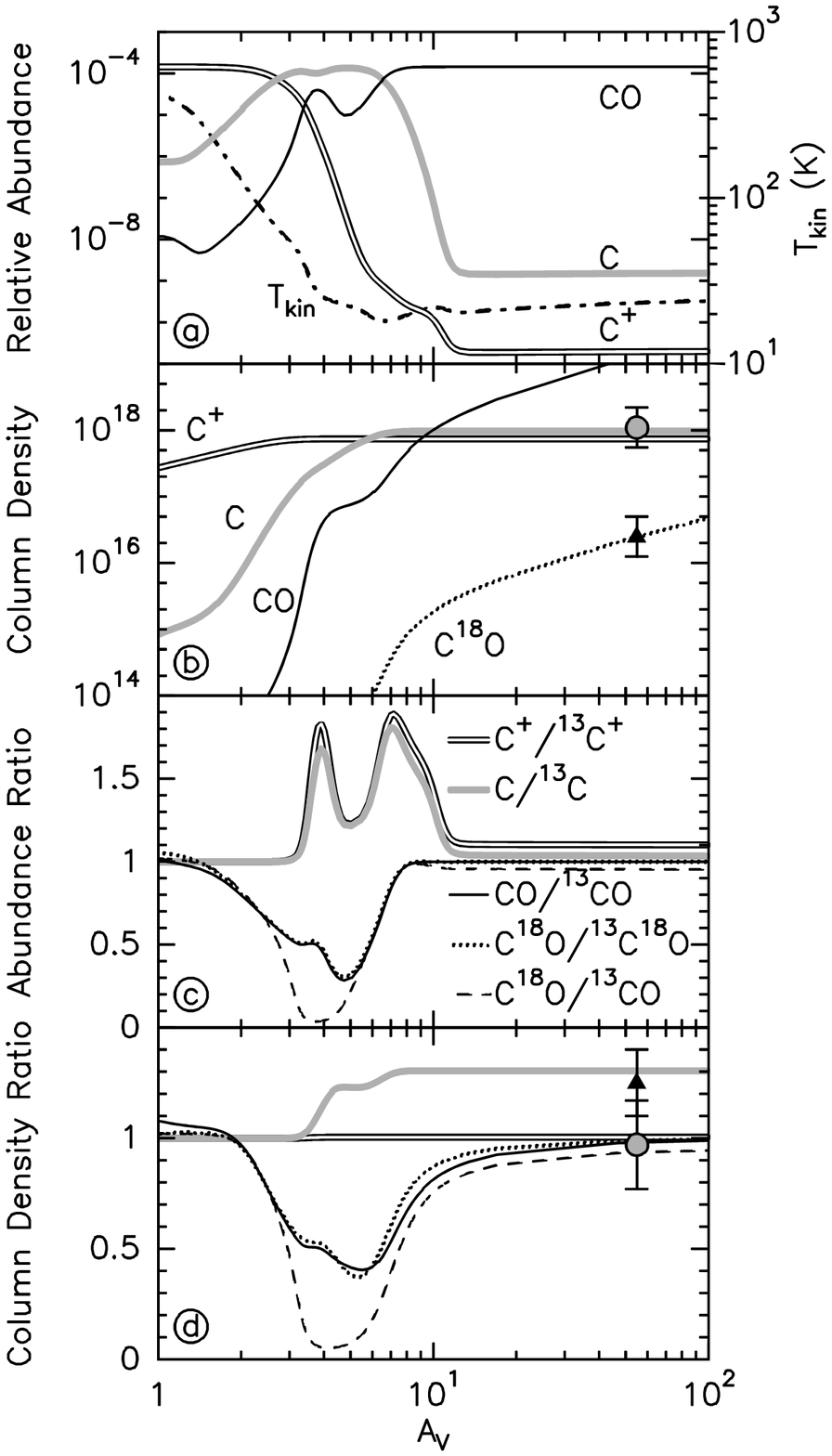}
\figcaption[keene_figure2.eps]{\small Output of the PDR model.  {\it a})
  The abundance of the major carbon species as a function of $A_V$
  into the cloud. {\it b}) The column density integrated from the
  cloud surface to the visual extinction on the $A_V$ axis. The
  markers show the observed column densities: filled gray circle --
  $N({\rm C}) = 1.1\times 10^{18}$ \protect\percmsq, black triangle --
  $N(\protect\mCeiO) = 2.5\times 10^{16}$ \protect\percmsq. {\it c}) The local
  abundance ratio of \protect\twC\ isotopes or isotopomers to \protect\thC\ ones,
  normalized by their input ratios.  {\it d}) The column density
  ratios integrated from the cloud surface to the visual extinction on
  the $A_V$ axis with the same key as shown in {\it c}. The markers
  show the observed column density ratios: filled gray circle --
  $N(\protect\mtwC) /60\; N(\protect\mthC) = 0.97$, 
  black triangle -- $N(\protect\mCeiO) /60\;
  N(\protect\mthCeiO) = 1.25$.\label{model}}
\vspace{0.3in}

\acknowledgments

The CSO is supported by the NSF under contract AST 96-15025. We thank
the staff of the CSO for their support, Richard Chamberlin and
Maryvonne Gerin for help in the observations, Jacques Le Bourlot for
help with the PDR model, Bill Langer and Ewine van Dishoeck for
reading and commenting on the manuscript, and Henrik Klein
and Frank Lewen for sharing their data before publication.

\end{document}